\documentclass[journal]{IEEEtran}
\usepackage{cite}

\usepackage{subfigure}
\ifCLASSINFOpdf
\usepackage[pdftex]{graphicx}
  \graphicspath{{../pdf/}{../jpeg/}}
  \DeclareGraphicsExtensions{.pdf,.jpeg,.png}
\else
  \usepackage[dvips]{graphicx}
  \graphicspath{{../eps/}}
  \DeclareGraphicsExtensions{.eps}
\fi
\usepackage{amsmath}
\usepackage{cases}
\usepackage{amsfonts}
\usepackage{subeqnarray}
\usepackage[ruled,linesnumbered]{algorithm2e}
\makeatletter
\newcommand{\nosemic}{\renewcommand{\@endalgocfline}{\relax}}
\newcommand{\dosemic}{\renewcommand{\@endalgocfline}{\algocf@endline}}
\let\oldnl\nl
\newcommand{\nonl}{\renewcommand{\nl}{\let\nl\oldnl}}
\makeatother
\usepackage[export]{adjustbox} 
\usepackage{booktabs}
\usepackage{setspace}
\usepackage{xcolor}
\usepackage{mathrsfs}
\usepackage{amsmath}
\usepackage{array}
\usepackage{amssymb}
\usepackage{amsthm}
\usepackage{microtype}

\allowdisplaybreaks
\begin{document}
\setstretch{0.93}
\title{\textls[-25]{Privacy-preserving Distributed Probabilistic Load Flow}}
\author{Mengshuo~Jia, 
		Yi~Wang,
Chen~Shen 
      and Gabriela~Hug

\thanks{M. Jia and C. Shen are with the State Key Laboratory of Power Systems, Tsinghua University, 100084 Beijing, China. Y. Wang and G. Hug are with the Power Systems Laboratory, ETH Zurich, 8092 Zurich, Switzerland.}}
        

\maketitle

\begin{abstract}
Probabilistic load flow (PLF) allows to evaluate uncertainties introduced by renewable energy sources on system operation. 
Ideally, the PLF calculation is implemented for an entire grid requiring all the parameters of the transmission lines and node load/generation to be available. However, in a multi-regional interconnected grid, the independent system operators (ISOs) across regions may not share the parameters of their respective areas with other ISOs. Consequently, the challenge is how to identify the functional relationship between the flows in the regional grid and the uncertain power injections of renewable generation sources across regions without full information about the entire grid. To overcome this challenge, we first propose a privacy-preserving distributed accelerated projection-based consensus algorithm for each ISO to calculate the corresponding coefficient matrix of the desired functional relationship. Then, we leverage a privacy-preserving accelerated average consensus algorithm to allow each ISO to obtain the corresponding constant vector of the same relationship. Using the two algorithms, we finally derive a privacy-preserving distributed PLF method for each ISO to analytically obtain its regional joint PLF in a fully distributed manner without revealing its parameters to other ISOs. The correctness, effectiveness, and efficiency of the proposed method are verified through a case study on the IEEE 118-bus system.
\end{abstract}

\IEEEpeerreviewmaketitle


\section{Introduction}

\IEEEPARstart{T}{he} growing penetration of renewable generation leads to an increase in uncertainties in power system operation. Probabilistic load flow (PLF) methods can effectively evaluate the underlying operational risks \cite{PRUSTY20171286}. However, in a multi-regional interconnected power system, multiple independent system operators (ISOs) perform regional operation \cite{982193}. Given the multi-regional interconnections and output coupling of renewable energy generation, the PLF of any region cannot be calculated individually but should take into account the uncertainties introduced in the other regions. Knowledge of system topology and parameters therefore is a prerequisite for the regional PLF calculation. However, each regional ISO only has access to the parameters within its area and usually may not share its system information with other ISOs \cite{7962302}. In this case, a distributed approach can provide the means to obtain the desired regional PLFs without sharing system data. 

To date, a number of methods have been proposed for centralized PLF calculation and can be divided into three categories: numerical, approximate, and analytical methods. Numerical methods rely on massive-scenario load flow calculation to extract the corresponding statistics. Although load flow model selection in numerical methods is usually AC-based, linear models are also used to reduce the computational burden \cite{CARPINELLI2015283}. The most common numerical method is Monte Carlo simulation \cite{CARMONADELGADO2015266,XIAO2018677,CARPINELLI2015283}. To improve the efficiency of numerical methods, importance sampling \cite{8462798}, Latin hypercube sampling \cite{LIU2016354}, and simple random sampling \cite{8049316} have been adopted. Approximate methods on the other hand aim to estimate the statistics of load flows using limited samples from known locations. Selecting appropriate samples that keep sufficient uncertain information of random power injections determines the success of these methods \cite{6192342}. Generally, approximate methods are also AC-based. The most frequently used approximate method is the point estimation method, including the two-point \cite{1525114}, multi-point \cite{4349110}, and fast \cite{7274097} schemes. As the unscented transformation can improve performance when propagating the mean and covariance information through nonlinear operations \cite{1271397}, it has been widely used to develop new approximate methods \cite{6192342,7934487}. Lastly, analytical methods aim to convert the probability distribution of the random power injections into PLF using functional relationships. However, as the AC model is highly nonlinear with implicit solutions, the explicit relationship may be inaccessible. Thus, analytical methods are mainly based on approximate load flow models, such as linear \cite{7784832}, generalized polynomial chaos \cite{8676044}, and low-rank approximation models \cite{8630065}. Among them, the linear model is the most common and serves as foundation for various analytical methods, including the convolution method \cite{7784832}, cumulant method with series expansion \cite{VILLANUEVA20141}, and a method based on Gaussian mixture model (GMM) \cite{7574307}. 

In the abovementioned centralized methods for PLF calculation, regional ISOs should share their parameters with the other ISOs to form the model of the entire grid. To avoid the need for this complete model, a distributed calculation strategy can be adopted. Some works have been conducted on distributed and deterministic load flow calculation, e.g.,  \cite{6987287,7962302} and \cite{4517082}. In this paper, we take a further step by proposing a privacy-preserving distributed PLF method. The distributed calculation only requires communication between adjacent ISOs, omitting centralized data collection. Moreover, privacy preservation is achieved as follows: 1) each ISO only needs its own parameters for calculation; 2) no ISO can deduce the parameters of others using the communicated information; and 3) no ISO can acquire the PLF of other regions.

For proposing the privacy-preserving distributed PLF method, we choose a specific PLF algorithm. Specifically, we first select the state-independent voltage-angle decoupled linearized power flow (DLPF) model \cite{7782382} as the load flow model, whose performance has been verified in \cite{8269410,8413105}. Then, to consider the correlations among renewable energy sources and variable load flows, we use GMM as the probability model to accurately represent multi-dimensional random variables subject to arbitrary distributions \cite{5298967}. Further, the GMM-based PLF method presented in \cite{7574307} is also adopted. Next, we combine the DLPF model and the GMM-based PLF method to obtain a base PLF method, which allows to analytically convert the probability distribution of the random injections into the joint PLF. The key to this base PLF method lies in finding the functional relationship between load flow in each region and random injected power over the entire grid. 

We further modify the base PLF method to obtain its privacy-preserving distributed version. To this end, we first reformulate the centralized PLF calculation into a distributed framework for multiple ISOs. The distributed framework for each ISO consists of two parts: 1) calculation of the coefficient matrix in the functional relationship between its regional load flow and random injected power over the whole grid, and 2) calculation of the constant vector in the same relationship. For the first part, we propose a privacy-preserving distributed accelerated projection-based consensus (APC) algorithm. This algorithm can enable each ISO to obtain the coefficient matrix through local calculations and privacy-preserving neighboring communication. For the second part, we leverage the privacy-preserving accelerated average consensus (AAC) algorithm in \cite{jia2020privacy} for each ISO to obtain the corresponding constant vector in a distributed and privacy-preserving manner. Based on these two algorithms, we derive the proposed privacy-preserving distributed PLF method. To the best of our knowledge, this is the first privacy-preserving distributed method for PLF calculation. 

The contributions of this paper can be summarized as follows:
\begin{itemize}
	\item We derive a distributed PLF framework for multiple ISOs. This framework unveils the requirements for a  privacy-preserving distributed PLF method.
	\item We propose a privacy-preserving distributed PLF method. This method enables every ISO to analytically obtain only its own regional joint PLF via a fully distributed manner without revealing its parameters to other ISOs. Meanwhile, this method includes a novel privacy-preserving distributed APC algorithm.
\end{itemize}

The rest of this paper is organized as follows. In Section II, we revisit the centralized PLF framework, which is then reformulated into a distributed version. In Section III, the privacy-preserving distributed APC algorithm is developed. Section IV describes the privacy-preserving AAC algorithm. Based on these two algorithms, the proposed privacy-preserving distributed PLF method is derived in Section V. Case studies are performed in Section VI. Section VII concludes this paper.

\section{Problem Formulation}
In this section, we first revisit the centralized GMM-based PLF method with the DLPF model. Then, the privacy issues of the centralized PLF framework are revealed. To mitigate these issues, we formulate the distributed PLF framework. 

\subsection{Centralized PLF Framework}

As the load flow model of the centralized PLF, the DLPF model assumes that $\cos\theta_{ij}\approx1$ and neglects $(V_i-1)^2$ and $(V_i-1)(V_j-1)$, because they are two orders of magnitude smaller than $V_i$ and $V_j$, respectively. Under these assumptions, the basic formulation of the DLPF model can be represented by \cite{7782382}
\begin{equation} 
\label{basic}     
\boldsymbol{A} \boldsymbol{x} = \boldsymbol{b},
\end{equation}

\noindent where 

\begin{align}
\boldsymbol{A} & = -
\left[                 
  \begin{array}{ccc}   
    \boldsymbol{B}^\prime_{\mathcal{SS}} \!\! & \!\! -\boldsymbol{B}^\prime_{\mathcal{SL}} \!\!&\!\! -\boldsymbol{G}_{\mathcal{SL}} \\
    \boldsymbol{B}^\prime_{\mathcal{LS}} \!\!&\!\! -\boldsymbol{B}^\prime_{\mathcal{LL}} \!\!&\!\! -\boldsymbol{G}_{\mathcal{LL}} \\ 
    \boldsymbol{B}^\prime_{\mathcal{WS}} \!\!&\!\! -\boldsymbol{B}^\prime_{\mathcal{WL}} \!\!&\!\! -\boldsymbol{G}_{\mathcal{WL}} \\ 
    \boldsymbol{G}_{\mathcal{LS}} \!\!&\!\! \boldsymbol{G}_{\mathcal{LL}} \!\!&\!\! -\boldsymbol{B}_{\mathcal{LL}} \\
    \boldsymbol{G}_{\mathcal{WS}} \!\!&\!\! \boldsymbol{G}_{\mathcal{WL}} \!\!&\!\! -\boldsymbol{B}_{\mathcal{WL}} \\ 
  \end{array}
\right] \in \Re^{N\times N }, \label{A} \\
\boldsymbol{x} & = 
\left[                 
  \begin{array}{ccccc}   
    \boldsymbol{\theta}_{\mathcal{S}}^T &
    \boldsymbol{\theta}_{\mathcal{L}}^T & \boldsymbol{\theta}_{\mathcal{W}}^T &
    \boldsymbol{V}_{\mathcal{L}}^T &
    \boldsymbol{V}_{\mathcal{W}}^T
  \end{array} \label{x}
\right]^T \in \Re^{N},
\end{align}

\noindent and
\begin{align}
\boldsymbol{b} & = 
\left[                 
  \begin{array}{c}   
    \boldsymbol{P}_{\mathcal{S}} \\
    \boldsymbol{P}_{\mathcal{L}} \\ 
    \boldsymbol{P}_{\mathcal{W}} \\ 
    \boldsymbol{Q}_{\mathcal{L}} \\
    \boldsymbol{Q}_{\mathcal{W}} \\  
  \end{array}
\right] \!\!+\!\! 
\left[                 
  \begin{array}{ccc}   
    \boldsymbol{B}^\prime_{\mathcal{SR}} \!\! & \!\! -\boldsymbol{G}_{\mathcal{SR}} \!\!&\!\! -\boldsymbol{G}_{\mathcal{SS}} \\
    \boldsymbol{B}^\prime_{\mathcal{LR}} \!\!&\!\! -\boldsymbol{G}_{\mathcal{LR}} \!\!&\!\! -\boldsymbol{G}_{\mathcal{LS}} \\ 
    \boldsymbol{B}^\prime_{\mathcal{WR}} \!\!&\!\! -\boldsymbol{G}_{\mathcal{WR}} \!\!&\!\! -\boldsymbol{G}_{\mathcal{WS}} \\
    \boldsymbol{G}_{\mathcal{LR}} \!\!&\!\! -\boldsymbol{B}_{\mathcal{LR}} \!\!&\!\! -\boldsymbol{B}_{\mathcal{LS}} \\
    \boldsymbol{G}_{\mathcal{WR}} \!\!&\!\! -\boldsymbol{B}_{\mathcal{WR}} \!\!&\!\! -\boldsymbol{B}_{\mathcal{WS}} \\  
  \end{array}
\right]\!\!
\left[                 
  \begin{array}{c}   
    \boldsymbol{\theta}_{\mathcal{R}} \\
    \boldsymbol{V}_{\mathcal{R}} \\  
    \boldsymbol{V}_{\mathcal{S}}   
  \end{array}
\right] \notag \\
& = \left[                 
	  \begin{array}{ccccc}   
	    \widetilde{\boldsymbol{P}}_{\mathcal{S}}^T &
	    \widetilde{\boldsymbol{P}}_{\mathcal{L}}^T &  
	    \boldsymbol{P}_{\mathcal{W}}^T \!\!+\!\! \boldsymbol{\rho}_{\mathcal{W}}^T &  
	    \widetilde{\boldsymbol{Q}}_{\mathcal{L}}^T &
	    \boldsymbol{Q}_{\mathcal{W}}^T \!\!+\!\! \boldsymbol{\sigma}_{\mathcal{W}}^T
	  \end{array}
	\right]^T \!\!\! \in \Re^{N}. \label{b with wind}
\end{align}

In the above equations, $\boldsymbol{G}$ and $\boldsymbol{B}$ are the conductance and susceptance matrices, where superscript $\prime$ represents the matrix without shunt elements. Subscripts $\mathcal{R}$, $\mathcal{S}$, and $\mathcal{L}$ correspond to the $V\theta$, $PV$, and $PQ$ buses, respectively, while $\mathcal{W}$ corresponds to $M$ buses with random injected power. For example, $\boldsymbol{P}_{\mathcal{S}}$ consists of the given active power injections of $PV$ buses, while the given reactive power injections of $PQ$ buses are included in $\boldsymbol{Q}_{\mathcal{L}}$. For more details, please refer to \cite{7782382}.


Substituting \eqref{b with wind} into \eqref{basic}, we can obtain $\boldsymbol{x}$ as a linear function of random injected power values $\boldsymbol{P}_{\mathcal{W}}$ and $\boldsymbol{Q}_{\mathcal{W}}$:
\begin{align}
	\label{linear function}
	\boldsymbol{x} = \boldsymbol{\alpha}\boldsymbol{P}_{\mathcal{W}} + \boldsymbol{\beta}\boldsymbol{Q}_{\mathcal{W}} + \boldsymbol{\gamma},
\end{align}
\noindent where
\begin{align}
	\boldsymbol{\gamma} & =  
	\boldsymbol{A}^{-1}
	\left[                 
	  \begin{array}{ccccc}   
	    \widetilde{\boldsymbol{P}}_{\mathcal{S}}^T &
	    \widetilde{\boldsymbol{P}}_{\mathcal{L}}^T &  
	    \boldsymbol{\rho}_{\mathcal{W}}^T &  
	    \widetilde{\boldsymbol{Q}}_{\mathcal{L}}^T &
	    \boldsymbol{\sigma}_{\mathcal{W}}^T
	  \end{array}
	\right]^T  \in \Re^{N}. \label{gamma}
\end{align}
\noindent and $\boldsymbol{\alpha} \in \Re^{N \times M}$ and $\boldsymbol{\beta} \in \Re^{N \times M}$ consist of the elements in $\boldsymbol{A}^{-1}$ corresponding to $\boldsymbol{P}_{\mathcal{W}}$ and $\boldsymbol{Q}_{\mathcal{W}}$. 


After obtaining the functional relationship in \eqref{linear function}, we can analytically compute the joint PLF of each region using the GMM-based PLF method in \cite{7574307}. First, let $\boldsymbol{x}_i \in \Re^{N_i} $ be the vector that consists of the states of region $i$. Then, we have 
\begin{align}
	\label{xi}
	\boldsymbol{x}_i = \boldsymbol{\alpha}_i\boldsymbol{P}_{\mathcal{W}} + \boldsymbol{\beta}_i\boldsymbol{Q}_{\mathcal{W}} + \boldsymbol{\gamma}_i,
\end{align}
where $\boldsymbol{\alpha}_i \in \Re^{N_i \times M}$ is the submatrix of $\boldsymbol{\alpha}$ whose rows correspond to the states of region $i$, and $\boldsymbol{\beta}_i$ is analogously defined. In addition, $\boldsymbol{\gamma}_i$ is a subvector of $\boldsymbol{\gamma}$, whose elements also correspond to the states of region $i$. Note that $\boldsymbol{\alpha}_i$ and $\boldsymbol{\beta}_i$ are the coefficient matrices of the functional relationship in \eqref{xi}, while $\boldsymbol{\gamma}_i$ is the constant vector.

Second, denote the GMM-based joint probability distribution of $\boldsymbol{P}_{\mathcal{W}}$ and $\boldsymbol{Q}_{\mathcal{W}}$ as
\begin{align}
	f(\boldsymbol{P}_{\mathcal{W}},\boldsymbol{Q}_{\mathcal{W}}) = \sum\nolimits_{k=1}^K w_k \ 
	\mathcal N_k(\boldsymbol{P}_{\mathcal{W}},\boldsymbol{Q}_{\mathcal{W}}|\boldsymbol{\mu}_k, \boldsymbol{\Sigma}_k), \label{GMM}  
\end{align}
\noindent where $\mathcal N_k(\cdot)$ is the $k$-th $2M$-dimensional Gaussian distribution with mean $\boldsymbol{\mu}_k$ and covariance $\boldsymbol{\Sigma}_k$. The weighting coefficient of $\mathcal N_k(\cdot)$ is $w_k$. 

Finally, using the functional relationship in \eqref{xi} and the parameters in \eqref{GMM}, the joint probability distribution of $\boldsymbol{x}_i$ can be expressed as \cite{7574307}
\begin{align}
	g(\boldsymbol{x}_i) = \sum\nolimits_{k=1}^K w_k \ 
	\mathcal N_k(\boldsymbol{x}|\boldsymbol{\lambda}_{ki}, \boldsymbol{\Delta}_{ki}), \label{GMM xi}  
\end{align}
\noindent where
\begin{align}
	\boldsymbol{\lambda}_{ki} & = [\boldsymbol{\alpha}_i^T \  \boldsymbol{\beta}_i^T]^T\boldsymbol{\mu}_k + \boldsymbol{\gamma}_i \label{xi mean} \\\boldsymbol{\Delta}_{ki} & = [\boldsymbol{\alpha}_i^T \  \boldsymbol{\beta}_i^T]^T\boldsymbol{\Sigma}_k[\boldsymbol{\alpha}_i^T \  \boldsymbol{\beta}_i^T]. \label{xi covariance}
\end{align}

Note that $g(\boldsymbol{x}_i)$ is the joint probability distribution of all the states in region $i$ (i.e., joint PLF of this region).

\subsection{Privacy Issues of Centralized PLF Framework}

Based on the above PLF formulation, we know that once the functional relationship in \eqref{xi} and joint probability distribution in \eqref{GMM} are known, the joint PLF in \eqref{GMM xi} can be derived directly. 

Establishing the joint probability distribution in \eqref{GMM} requires historical data of the random injected power for training. In this paper, we assume that these data are publicly available, like in the case of electricity metadata generated from European renewable sources available at Eurostat. Meanwhile, the reactive power can be calculated from the active power by assuming a constant power factor \cite{8630065}. Therefore, each ISO can directly obtain $f(\boldsymbol{P}_{\mathcal{W}},\boldsymbol{Q}_{\mathcal{W}})$ using a method such as the expectation--maximization algorithm for training \cite{5298967}. 

Identifying the functional relationship in \eqref{xi} requires the model of the entire grid (i.e., complete $\boldsymbol{A}$ and $\boldsymbol{b}$), because $\boldsymbol{\alpha}_i$ and $\boldsymbol{\beta}_i$ consist of the elements in $\boldsymbol{A}^{-1}$. In addition, each element of  $\boldsymbol{\gamma}_i$ is the inner product of $\boldsymbol{b}$ and a row in $\boldsymbol{A}^{-1}$. However, each ISO does not have complete information on $\boldsymbol{A}$ and $\boldsymbol{b}$ but only accesses the following information:
\begin{itemize}
	\item parameters of transmission lines within its region
	\item parameters of tie-lines linked to its region
	\item load and generation information within its region
	\item states of buses within its region
	\item states of ends of tie-lines linked to its region
\end{itemize}

Using the available information, each ISO can only form submatrices of $\boldsymbol{A}$ and $\boldsymbol{b}$, that is, ISO $i$ can only form $\boldsymbol{A}_i \in \Re^{N_i \times N} $ and $\boldsymbol{b}_i \in \Re^{N_i} $, where $\boldsymbol{A}_i$ consists of the $N_i$ conductance and susceptance rows related to the buses within region $i$, and $\boldsymbol{b}_i$ consists of $N_i$ power injection values related to the same buses. If we consider $H$ ISOs, the relationships among the above submatrices are expressed as
\begin{align}
	\bigcup_{i=1}^H \boldsymbol{A}_i = \boldsymbol{A} \ \text{,} \ \bigcap_{i=1}^H \boldsymbol{A}_i = \varnothing\ \text{,} \ \bigcup_{i=1}^H \boldsymbol{b}_i = \boldsymbol{b} \ \text{,} \  \bigcap_{i=1}^H \boldsymbol{b}_i = \varnothing \notag.
\end{align}

In the centralized PLF framework, $\boldsymbol{A}_i$ and $\boldsymbol{b}_i $ of each ISO are collected to form the complete $\boldsymbol{A}$ and $\boldsymbol{b}$. This information sharing leads to privacy issues and may be refused by ISOs. 

For ISO $i$ to calculate its regional joint PLF while preserving privacy, we need to answer the following question: if ISO $i$ ($\forall i$) knows only $\boldsymbol{A}_i$ and $\boldsymbol{b}_i$, how can it obtain only $\boldsymbol{\alpha}_i$, $\boldsymbol{\beta}_i$, and $\boldsymbol{\gamma}_i$ in \eqref{xi}? Next, we will devise a distributed PLF framework to answer this question.

\subsection{Distributed Framework for Coefficient Matrix Calculation}

Before devising the distributed framework, we introduce a vector $\boldsymbol{P}_{\ell}$ to ensure that each ISO cannot obtain the PLF of other regions. First, ISO $i$ ($\forall i$) chooses an element $\widetilde{P}_i$ from $\widetilde{\boldsymbol{P}}_{\mathcal{L}}^T$ in \eqref{b with wind}, where $\widetilde{P}_i \in \Re$ is a nodal active power injection within region $i$ only available to ISO $i$. Next, we use these $H$ elements to form $\boldsymbol{P}_{\ell}$:
\begin{align}
	\boldsymbol{P}_{\ell}=
	\left[\widetilde{P}_1,...,\widetilde{P}_H
	\right]^T \in \widetilde{\boldsymbol{P}}_{\mathcal{L}}^T.
\end{align}

After that, we reformulate \eqref{linear function} into an augmented form by introducing $\boldsymbol{P}_{\ell}$:
\begin{align}
	\label{modified linear function}
	\boldsymbol{x} = \boldsymbol{\alpha}\boldsymbol{P}_{\mathcal{W}} + \boldsymbol{\beta}\boldsymbol{Q}_{\mathcal{W}} + \boldsymbol{\epsilon}\boldsymbol{P}_{\ell} + \boldsymbol{\gamma}^{\prime},
\end{align}
\noindent where 
\begin{align}
	\boldsymbol{\gamma} = \boldsymbol{\epsilon}\boldsymbol{P}_{\ell} + \boldsymbol{\gamma}^{\prime} \label{modified gamma}
\end{align}
\noindent and $\boldsymbol{\epsilon} \in \Re^{N \times H}$ consists of the elements in $\boldsymbol{A}^{-1}$ corresponding to $\boldsymbol{P}_{\ell}$. Note that Section IV will further discuss why introducing $\boldsymbol{P}_{\ell}$ can prevent ISOs from obtaining the PLF of other regions.

For ISO $i$ to obtain coefficient matrices $\boldsymbol{\alpha}_i$ and $\boldsymbol{\beta}_i$, we define $\boldsymbol{\Lambda}$ as follows:
\begin{align}
	\boldsymbol{\Lambda} =  
	\left[                 
	  \begin{array}{cccc}   
	    \boldsymbol{\alpha} &
	    \boldsymbol{\beta} & 
	    \boldsymbol{\epsilon} & 
	    \boldsymbol{\gamma}^{\prime}
	  \end{array}
	\right]  \in \Re^{N \times \hat{M}},\label{augmented Lambda}
\end{align}
where 
\begin{align}
	\hat{M} = 2M + H + 1.
\end{align}

ISO $i$ should extract $\boldsymbol{\alpha}_i$ and $\boldsymbol{\beta}_i$ from $\boldsymbol{\Lambda}$. To compute $\boldsymbol{\Lambda}$, the ISOs need to choose $\hat{M}$ publicly known observations of $\boldsymbol{P}_{\mathcal{W}}$ and $\boldsymbol{Q}_{\mathcal{W}}$, where $\boldsymbol{P}_{\mathcal{W}}(k) \in \Re^{M}$ is the $k$-th observation of $\boldsymbol{P}_{\mathcal{W}}$. Besides, the ISOs also need to generate $\hat{M}$ artificial and publicly known data segments of $\boldsymbol{P}_{\ell}$, where the $k$-th data segment is represented by $\boldsymbol{P}_{\ell}(k)$. Substituting $\boldsymbol{P}_{\mathcal{W}}(k)$, $\boldsymbol{Q}_{\mathcal{W}}(k)$, and $\boldsymbol{P}_{\ell}(k)$ into $\boldsymbol{b}$ generates $\boldsymbol{b}(k)$. Then, the following equation holds:
\begin{equation}
	\label{modified compute Lambda}
	\boldsymbol{\Lambda}\boldsymbol{\Pi} = \boldsymbol{A}^{-1}\boldsymbol{B},
\end{equation}
\noindent where
\begin{align}
	\boldsymbol{\Pi} &  =  
	\left[                 
	  \begin{array}{ccc}   
	    \boldsymbol{P}_{\mathcal{W}}(1) & \cdots & \boldsymbol{P}_{\mathcal{W}}(\hat{M}) \\
	    \boldsymbol{Q}_{\mathcal{W}}(1) & \cdots & \boldsymbol{Q}_{\mathcal{W}}(\hat{M})\\
	    \boldsymbol{P}_{\ell}(1) & \cdots & \boldsymbol{P}_{\ell}(\hat{M})\\  
	    1  & \cdots & 1 \\
	  \end{array}
	\right] \in \Re^{\hat{M} \times \hat{M}}, \label{modified Pi} \\
	\boldsymbol{B} &  =
	\left[
		\begin{array}{ccc}
			\boldsymbol{b}(1) & \cdots & \boldsymbol{b}(\hat{M})
		\end{array} 
	\right] \in \Re^{N \times \hat{M}}. \label{modified B}
\end{align}

Given that $\boldsymbol{\Pi}$ is available to all ISOs, if each ISO has the results of the right-hand side of \eqref{modified compute Lambda}, it could then compute $\boldsymbol{\Lambda}$. Computing the right-hand side of \eqref{modified compute Lambda} is essentially calculating $\boldsymbol{\mathcal{X}}\in \Re^{N \times \hat{M}}$ in 
\begin{equation}
	\boldsymbol{A}\boldsymbol{\mathcal{X}} =\boldsymbol{B}. \label{X} 
\end{equation}

However, similar to $\boldsymbol{A}$ and $\boldsymbol{b}$, ISO $i$ can only form a submatrix of $\boldsymbol{B}$, i.e., $\boldsymbol{B}_i \in \Re^{N_i \times \hat{M}} $, which consists of $N_i$ rows of injected power values related to the buses within region $i$. Therefore, ISO $i$ mathematically faces the problem of acquiring $\boldsymbol{\mathcal{X}}$ in 
\begin{align}
	\left[                 
	  \begin{array}{ccc}   
	    \boldsymbol{A}_{1}^T &
	    \cdots & 
	    \boldsymbol{A}_{H}^T  
	  \end{array}
	\right]^T
	\boldsymbol{\mathcal{X}} = 
	\left[                 
	  \begin{array}{ccc}   
	    \boldsymbol{B}_{1}^T &
	    \cdots & 
	    \boldsymbol{B}_{H}^T  
	  \end{array}
	\right]^T. \label{Xi}
\end{align}

\textbf{\textit{Remark 1}}: ISO $i$ ($\forall i$) calculating coefficient matrices $\boldsymbol{\alpha}_i$ and $\boldsymbol{\beta}_i$ is essentially solving \eqref{Xi} when only $\boldsymbol{A}_i$ and $\boldsymbol{B}_i$ are available. After solving \eqref{Xi}, ISO $i$ ($\forall i$) can then obtain $\boldsymbol{\Lambda}$ by \eqref{modified compute Lambda} and further extract $\boldsymbol{\alpha}_i$ and $\boldsymbol{\beta}_i$ from $\boldsymbol{\Lambda}$. Thus, a privacy-preserving distributed PLF method should guarantee that every ISO solves \eqref{Xi} in a privacy-preserving and fully distributed manner. 

\subsection{Distributed Framework for Constant Vector Calculation}

To allow ISO $i$ to obtain the constant vector $\boldsymbol{\gamma}_i$, we first define the index set of its states as $\boldsymbol{\Theta}_i$. Then, we define $\gamma_n$ as the $n$-th element of $\boldsymbol{\gamma}$. Clearly, $\boldsymbol{\gamma}_i$ consists of $\gamma_n$ ($n \in \boldsymbol{\Theta}_i$). Thus, based on \eqref{modified gamma}, $\gamma_n$ can be calculated as
\begin{align}
	\gamma_n  & = \gamma_n^{\prime} + H\psi_n, \ \ n \in \boldsymbol{\Theta}_i, \label{gamma n}  
\end{align}
where 
\begin{align}
	\psi_n & = \frac{1}{H} \sum\nolimits_{i=1}^H \epsilon_{ni}\widetilde{P}_i, \ \ n \in \boldsymbol{\Theta}_i. \label{psi n} 
\end{align}
In \eqref{gamma n}, $\gamma_n^{\prime}$ is the $n$-th element of $\boldsymbol{\gamma}^{\prime}$ and $\epsilon_{ni}$ is the element in row $n$ and column $i$ in $\boldsymbol{\epsilon}$. Note that both $\boldsymbol{\gamma}^{\prime}$ and $\boldsymbol{\epsilon}$ are known by all ISOs after they solved \eqref{Xi} and further obtain $\boldsymbol{\Lambda}$ in \eqref{modified compute Lambda}. Thus, if ISO $i$ has $\psi_n$ ($n \in \boldsymbol{\Theta}_i$) in \eqref{psi n}, it can then compute $\boldsymbol{\gamma}_i$ by \eqref{gamma n}. However, ISO $i$ only knows $\widetilde{P}_i$. In this case, ISO $i$ needs to solve the problem of how  to acquire $\psi_n$ ($n \in \boldsymbol{\Theta}_i$) when it only knows $\widetilde{P}_i$.

\textbf{\textit{Remark 2}}: Calculating the constant vector $\boldsymbol{\gamma}_i$ for ISO $i$ ($\forall i$) is essentially computing $\psi_n$ ($n \in \boldsymbol{\Theta}_i$) in \eqref{psi n} when the ISO only knows $\widetilde{P}_i$. Thus, a privacy-preserving distributed PLF method should enable each ISO to calculate \eqref{psi n} in a privacy-preserving and fully distributed manner. 

\section{Privacy-Preserving Distributed APC}

To enable every ISO to solve \eqref{Xi} in a privacy-preserving and fully distributed manner, we propose a privacy-preserving and fully distributed APC algorithm derived from the conventional APC algorithm \cite{8462630}. 

\subsection{APC Algorithm}

The APC algorithm \cite{8462630} aims to solve a system of linear equations that are partitioned such that each party only accesses a disjoint subset of the full set of equations and variables. In our case, the party is the ISO, and the system of linear equations is \eqref{Xi}. To obtain $\boldsymbol{\mathcal{X}}$, ISO $i$ first finds an initial solution $\boldsymbol{X}_i(0) \in \Re^{N \times \hat{M}}$ of $\boldsymbol{A}_i\boldsymbol{X}_i(0)=\boldsymbol{B}_i$ among infinitely many solutions. Then, ISO $i$ updates its initial solution via the APC algorithm as follows: 
\begin{equation}
	\boldsymbol{X}_i(t+1) = \boldsymbol{X}_i(t) + \varphi\boldsymbol{\Gamma}_i \left[ \boldsymbol{\overline{X}}(t)- \boldsymbol{X}_i(t)\right], \label{APC}
\end{equation}
\noindent where $\boldsymbol{\Gamma}_i\in \Re^{N \times N}$ is the projection matrix onto the nullspace of $\boldsymbol{A}_i$, as given in:
\begin{align}
	\boldsymbol{\Gamma}_i = \boldsymbol{I}-\boldsymbol{A}_i^T(\boldsymbol{A}_i \boldsymbol{A}_i^T)^{-1}\boldsymbol{A}_i  = \boldsymbol{I} - \boldsymbol{\Phi}_i, \label{Gamma}
\end{align}
\noindent $\boldsymbol{\overline{X}}(t)\in \Re^{N \times \hat{M}} $ is the estimation of the global solution $\boldsymbol{\mathcal{X}}$ at iteration $t$, as given in :
\begin{align}
\label{average 1} 
	\begin{split}
	& \boldsymbol{\overline{X}}(t) = (1-\eta)\boldsymbol{\overline{X}}(t-1) + \frac{\eta}{H}\sum\nolimits_{i=1}^H\boldsymbol{X}_i(t) \\
	& \boldsymbol{\overline{X}}(0) = \frac{1}{H}\sum\nolimits_{i=1}^H\boldsymbol{X}_i(0), 
	\end{split}
\end{align}
\noindent and $\boldsymbol{I}$ is the $N$-dimensional identity matrix. Besides, the optimal parameters of $\varphi$ and $\eta$ in \eqref{APC} and \eqref{average 1} are the solutions of the following equations:
\begin{equation}
	\label{varphi and eta}
	\left\{
	\begin{split}
	& \upsilon_{max}\, \varphi \,  \eta = (1+\sqrt{(\varphi-1)(\eta-1)})^2, \\
	& \upsilon_{min}\, \varphi \,  \eta = (1-\sqrt{(\varphi-1)(\eta-1)})^2, \\
	\end{split} \right.
\end{equation}
\noindent where $\upsilon_{max}$ and $\upsilon_{min}$ are the maximal and minimal eigenvalues of $\boldsymbol{\Upsilon} \in \Re^{N \times N} $:
\begin{align}
	\boldsymbol{\Upsilon} = \frac{1}{H} \sum\nolimits_{i=1}^H \boldsymbol{\Phi}_i.  \label{average 2}
\end{align}

Using the iterative process defined by \eqref{APC} - \eqref{average 2}, $\boldsymbol{X}_i$ ($\forall i$) converges to the global solution $\boldsymbol{\mathcal{X}}$ with the convergence rate \cite{8462630}:
\begin{align}
	r = 1-\frac{2}{\upsilon_{max}}\,\upsilon_{min}.
\end{align}

Note that computing \eqref{average 1} and \eqref{average 2} requires average calculations among all ISOs. For clarity, we summarize these average calculations as 
\begin{align}
	\boldsymbol{G} = \frac{1}{H} \sum\nolimits_{i=1}^H \boldsymbol{L}_i, \label{average 3}
\end{align}
where $\boldsymbol{L}_i $ represents $\boldsymbol{X}_i(t)$ or $\boldsymbol{\Phi}_i$, while $\boldsymbol{G}$ represents $\boldsymbol{\overline{X}}(t)$ or $\boldsymbol{\Upsilon}$ correspondingly. To compute \eqref{average 3}, the authors in \cite{8462630} use a center for data collection, calculation, and broadcasting. Once \eqref{average 3} is obtained by each ISO, other calculations of the APC algorithm can be performed independently.

\subsection{Privacy-Preserving Distributed APC Algorithm}
To develop a privacy-preserving and fully distributed APC algorithm without centralized data collection, each ISO needs to be able to compute \eqref{average 3} by local calculations and privacy-preserving communication with its neighbors. To this end, we use the privacy-preserving AAC algorithm proposed in \cite{jia2020privacy}. 

The privacy-preserving AAC algorithm is based on graph theory. Specifically, all ISOs should form a connected and publicly available graph consisting of $H$ nodes and some edges, where each edge between a pair of nodes represents bidirectional noiseless communication between the two corresponding ISOs. The neighborhood of ISO $i$, denoted by $\boldsymbol{\Omega}_i$, is defined as an index set of ISOs directly connected to ISO $i$. Meanwhile, the degree of ISO $i$ is represented by $d_i$.  The graph should guarantee that if $j \in \boldsymbol{\Omega}_i$, then $\boldsymbol{\Omega}_j \not\subset  \boldsymbol{\Omega}_i$, as described in \cite{jia2020privacy}. Under this graph, each ISO computes the elements in the so-called Metropolis weight matrix as follows:
\begin{equation}
	\label{weight matrix}
	W_{ij} =\left\{
	\begin{split}
	& \frac{1}{1+\max\{d_i,d_j\} }\quad \text{if $j \in \boldsymbol{\Omega}_i $} \\
	& 1-\sum\nolimits_{j \in \boldsymbol{\Omega}_i} W_{i,j} \quad \text{if $i=j$} \\
	& 0 \qquad\qquad\quad\quad\quad\ \  \text{Otherwise} \\
	\end{split} \right.
\end{equation}

Using the Metropolis weight matrix $\boldsymbol{W}\in \Re^{H \times H} $, each ISO further computes the accelerated Metropolis weight matrix $\boldsymbol{W}^\ast \in \Re^{H \times H} $ as follows:
\begin{align}
	\boldsymbol{W}^\ast \triangleq  (1+\varepsilon)  \boldsymbol{W} - \varepsilon \boldsymbol{I}, \label{W star}
\end{align}
\noindent where $\boldsymbol{I}$ is the $H$-dimensional identity matrix, and 
\begin{align}
	\varepsilon = \frac{\iota_{min}+\iota_2}{2-\iota_{min}-\iota_2}
\end{align}
is the optimal parameter for acceleration. Moreover, $\iota_{min}$ is the minimal eigenvalue of $\boldsymbol{W}$, and $\iota_{2}$ is its second largest eigenvalue.

After obtaining $\boldsymbol{W}^\ast$, ISO $i$ sets $\boldsymbol{y}_i(0)=\boldsymbol{L}_i$ and updates $\boldsymbol{y}_i(0)$ through the privacy-preserving AAC algorithm in 
\begin{equation}
	\boldsymbol{y}_i(t+1) = W_{i,i}^\ast \boldsymbol{y}_i^+(t) + \sum\nolimits_{j \in \boldsymbol{\Omega}_i } W_{i,j}^\ast \boldsymbol{y}_j^+ (t),  \label{AAC}
\end{equation}
\noindent where $\boldsymbol{y}_i^+(t)$ represents the true value of $\boldsymbol{y}_i(t)$ plus some noise:
\begin{equation}
	\label{mask}
	\begin{split}
		& \boldsymbol{y}_i^+(t) = \boldsymbol{y}_i(t) + \delta_i(t) - \delta_i(t-1), 
	\end{split}
\end{equation}

\noindent and noise
$\delta_i(t)$ is randomly selected from $[-\frac{\varrho}{2}\varsigma^{t+1}, \frac{\varrho}{2}\varsigma^{t+1}]$ by ISO $i$ with $\varrho>0$ and $\varsigma \in [0,1)$. By the iterative process, $\boldsymbol{y}_i$ converges to the average value of $\boldsymbol{L}_i$ for $i=1,...,H$ as follows:
\begin{equation}
	\lim\limits_{t \to \infty }{\boldsymbol{y}_i(t)} = \frac{1}{H} \sum\nolimits_{i=1}^M \boldsymbol{L}_i. \label{mean} 
\end{equation}
For a detailed proof, please refer to \cite{jia2020privacy}.

Based on the privacy-preserving AAC algorithm, we derive the privacy-preserving and fully distributed APC algorithm, which is detailed in Algorithm \ref{PPD APC}.
\begin{algorithm}
	\label{PPD APC}
	\caption{Privacy-preserving Distributed APC Algorithm for ISO $i$ ($\forall i$)}
	\KwIn{$\boldsymbol{A}_{i}$ and $\boldsymbol{B}_{i}$.}
    \KwOut{Solution $\boldsymbol{\mathcal{X}}$ of \eqref{Xi}}
	$t=0$\;
    Compute $\boldsymbol{X}_i(t)$ from $\boldsymbol{A}_i\boldsymbol{X}_i(t)=\boldsymbol{B}_i$\;
    Obtain $\boldsymbol{\overline{X}}(t)$ in \eqref{average 1} and $\boldsymbol{\Upsilon}$ in \eqref{average 2} by \eqref{AAC}-\eqref{mean}\;
    Compute $\boldsymbol{\Gamma}_i$ by \eqref{Gamma} and $\varphi$, $\eta$ by \eqref{varphi and eta}\;
    $t=1$\;
    \While{APC convergence criterion is not met}
	{
		
		Update $\boldsymbol{X}_i(t)$ by \eqref{APC}\;
		Obtain $\boldsymbol{\overline{X}}(t)$ in \eqref{average 1} by \eqref{AAC}-\eqref{mean}\;
		$t=t+1$\;
	}   
	Return $\boldsymbol{\mathcal{X}}=\boldsymbol{X}_i(t)$\;	
\end{algorithm}

It should be emphasized that in the privacy-preserving distributed APC algorithm, each ISO only needs local calculations (steps 2, 4, 7, and 11) and neighboring communications (steps 3 and 8). Thus, the proposed algorithm is fully distributed and no center for data collection is required. Meanwhile, the only information that ISO $i$ shares with its neighbors is $\boldsymbol{y}_i^+(t)$, which is masked by random noise. Therefore, the neighbors cannot deduce any private information from $\boldsymbol{y}_i^+(t)$, resulting in strict privacy protection despite communication.

\section{Privacy-Preserving AAC with Fake Input}

To enable ISO $i$ ($\forall i$) to obtain $\psi_n$ in \eqref{psi n} via a privacy-preserving and fully distributed fashion, we note that \eqref{psi n} is mathematically equivalent to \eqref{average 3}. Thus, ISO $i$ ($\forall i$) can still use the privacy-preserving AAC algorithm in \eqref{AAC} to calculate \eqref{psi n}.

However, to guarantee that ISO $i$ only obtains $\gamma_n$ ($n \in \boldsymbol{\Theta}_i$) after performing the privacy-preserving AAC algorithm, we introduce a fake value $\hat{P}_i$ here. The fake value $\hat{P}_i$ is randomly generated and only available to ISO $i$. Using this fake value, ISO $i$ sets $y_{ni}(0)$ ($n=1,...,N$) as follows:
\begin{equation}
	\label{fake initial}
	y_{ni}(0) =\left\{
	\begin{split}
	& \epsilon_{ni}\widetilde{P}_i\ \text{,}\ \ \  n \notin \boldsymbol{\Theta}_i \\
	& \epsilon_{ni}\hat{P}_i\ \text{,}\ \ \  n \in \boldsymbol{\Theta}_i \\
	\end{split} \right.
\end{equation}
\noindent where $y_{ni}(0)$ is the $n$-th element of $\boldsymbol{y}_i(0)$ in \eqref{AAC}. After performing the privacy-preserving AAC algorithm, $y_{ni}(t)$ converges to the following values for $n=1,...,N$:
\begin{equation}
	\label{fake mean}
	\lim\limits_{t \to \infty }{y_{ni}(t)} =\left\{
	\begin{split}
	& \frac{1}{H} \epsilon_{nj}\hat{P}_j + \frac{1}{H}\sum\nolimits_{k=1,k\neq j}^H \epsilon_{nk}\widetilde{P}_k \text{,}\ \ n \in \boldsymbol{\Theta}_j \\
	& \frac{1}{H} \epsilon_{ni}\hat{P}_i + \frac{1}{H}\sum\nolimits_{k=1,k\neq i}^H \epsilon_{nk}\widetilde{P}_k \text{,}\ \ n \in \boldsymbol{\Theta}_i \\
	\end{split} \right.
\end{equation}
\noindent where $\hat{P}_j$ is the random fake value chosen by any ISO $j$ ($j\neq i$). Clearly, ISO $i$ can compute the real $\psi_n$ ($n \in \boldsymbol{\Theta}_i$) from \eqref{fake mean} as
\begin{align}
	\psi_n = \lim\limits_{t \to \infty }{y_{ni}(t)}  - \frac{1}{H} \epsilon_{ni}\hat{P}_i + \frac{1}{H} \epsilon_{ni}\widetilde{P}_i. \label{real psi}
\end{align}
Then, ISO $i$ acquires $\gamma_n$ ($n \in \boldsymbol{\Theta}_i$) using \eqref{gamma n} (i.e., $\boldsymbol{\gamma}_i$). However, as ISO $i$ does not know $\hat{P}_j$ and $\widetilde{P}_j$ of ISO $j$, it cannot deduce $\psi_n$ ($n \in \boldsymbol{\Theta}_j$) from \eqref{fake mean}. Therefore, every $\gamma_n$ ($n \in \boldsymbol{\Theta}_j$) remains unknown to ISO $i$, disabling it to derive the PLF of another region.

\section{Privacy-Preserving Distributed PLF Method}
Based on the privacy-preserving distributed APC and privacy-preserving AAC algorithms, we propose the privacy-preserving distributed PLF method in Algorithm \ref{PPD PLF}.

\begin{algorithm}
	\label{PPD PLF}
	\caption{Privacy-Preserving Distributed PLF Method for ISO $i$ ($\forall i$)}
	\KwIn{$\boldsymbol{A}_{i}$ and $\boldsymbol{b}_{i}$.}
    \KwOut{Probability distribution $g(\boldsymbol{x}_i)$.}  
    Form $\boldsymbol{A}_i$ and $\boldsymbol{B}_i$ in \eqref{Xi}\;
    Perform Algorithm 1 to obtain $\boldsymbol{A}^{-1}\boldsymbol{B}$\;
    Acquire $\boldsymbol{\Lambda}$ in \eqref{modified compute Lambda}\;
    Choose secret fake value $\hat{P}_i$ randomly\;
    Set $y_{ni}(0)$ ($n=1,...,N$) according to \eqref{fake initial} to get $\boldsymbol{y}_i(0)$\;
    Perform privacy-preserving AAC algorithm\;
    Obtain results in \eqref{fake mean} after convergence\;
    Compute $\psi_n$ ($n \in \boldsymbol{\Theta}_i$) by \eqref{real psi}\;
    Form $\boldsymbol{\gamma}_i$ by \eqref{gamma n}\;
    Extract $\boldsymbol{\alpha}_i$ and $\boldsymbol{\beta}_i$ from $\boldsymbol{\Lambda}$\;
    Build $f(\boldsymbol{P}_{\mathcal{W}},\boldsymbol{Q}_{\mathcal{W}})$ in \eqref{GMM}\;
    Derive $g(\boldsymbol{x}_i)$ in \eqref{GMM xi} using $\boldsymbol{\alpha}_i$, $\boldsymbol{\beta}_i$ and $\boldsymbol{\gamma}_i$\;
\end{algorithm}

There are two points that should be noted about the proposed algorithm. First, except for steps 2 and 7 in Algorithm 2, the other steps are local calculations that can be conducted by each ISO. In addition, steps 2 and 7 are privacy-preserving distributed calculations. Therefore, Algorithm 2 provides a privacy-preserving distributed method that allows each ISO to only obtain the PLF of its own region through local calculations and privacy-preserving communication with its neighbors. No ISO can deduce private information of the other ISOs. Second, $g(\boldsymbol{x}_i)$ is a joint probability distribution that characterizes the stochastic features of all states in region $i$ considering their correlations. This distribution can provide a simultaneous and exact evaluation for the probability of multiple states being out of bound \cite{7574307}. Moreover, deriving the marginal or conditional probability distribution of a single state from $g(\boldsymbol{x}_i)$ is also straightforward \cite{WANG2018771}.

\section{Case Study}

\subsection{Settings}
\begin{figure}[h]
	\centering  
	\includegraphics[width=3.5in,center]{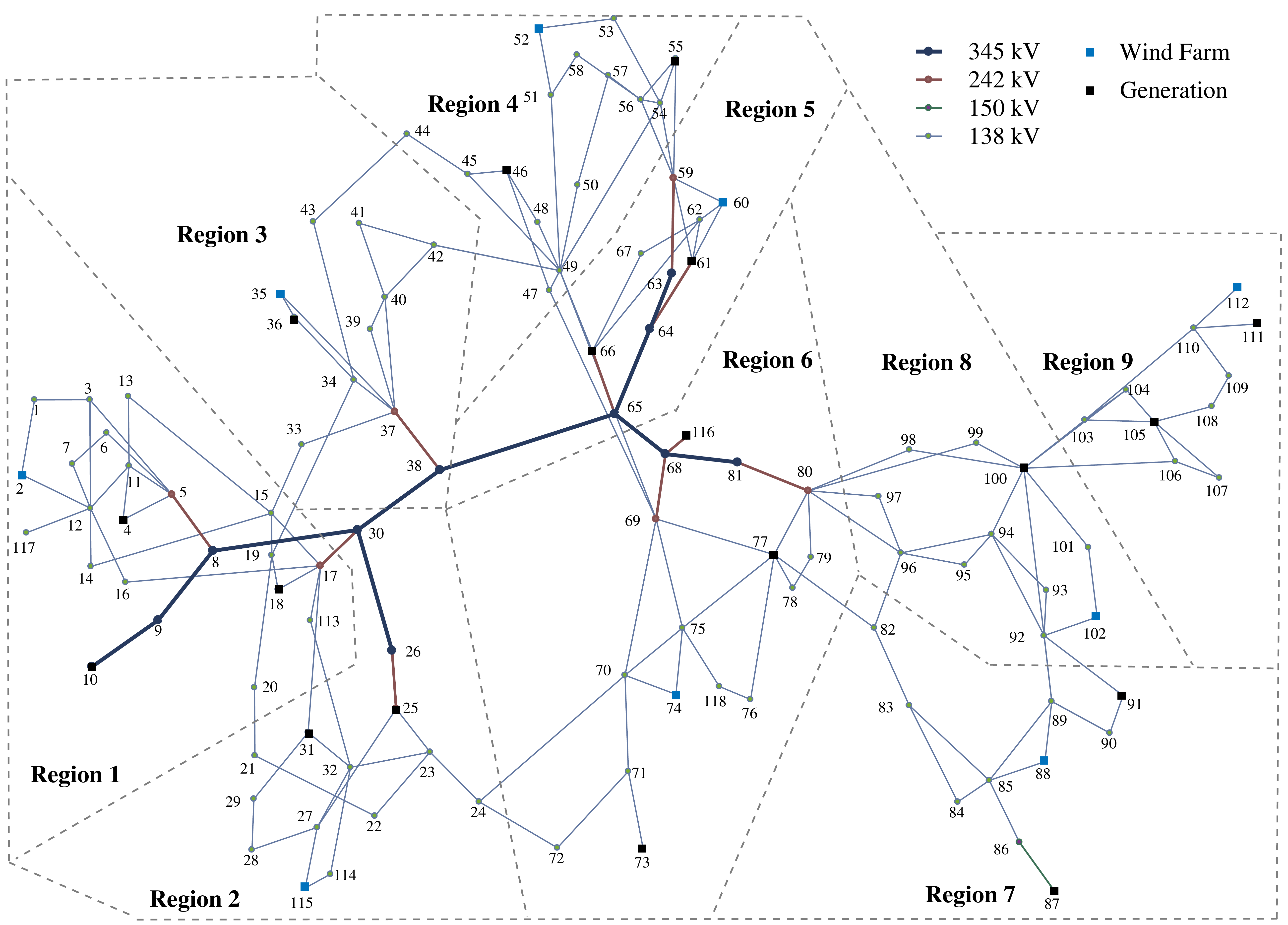}
	\caption{multi-regional interconnected grid based on IEEE 118-bus system.}
	\label{case_grid}
\end{figure}

We modified the IEEE 118-bus system for conducting a case study. We randomly divided the system into nine regions, as detailed in Fig. \ref{case_grid}. Note that a wind farm was added to each region to represent the random power injections. In addition, we used the data from the Eastern Wind Integration Dataset published by the US National Renewable Energy Laboratory to simulate historical data of wind farms. 

Besides, under different levels of power injections, the power flow results obtained from the DLPF model show small and approximate constant deviations compared to the real AC results \cite{8289421}. So for better performance, one can estimate the deviations using a given set of power injections and then add the deviations to the corresponding results of the DLPF model to obtain a complemented version. In the following case studies, all the evaluated methods that embed the DLPF model use the complemented version.

Moreover, all the experiments were coded in MATLAB and run on an i5-7267U 3.1 GHz processor with 8 GB RAM. 

\subsection{Correctness Verification}
In this paper, correctness means that the calculation results of the distributed algorithm and its corresponding centralized algorithm should be consistent. 

To verify the correctness of the proposed privacy-preserving distributed APC algorithm, we first computed $\boldsymbol{\mathcal{X}}_{real}=\boldsymbol{A}^{-1}\boldsymbol{B}$ in \eqref{X} in a centralized way and used the result as benchmark. Then, we used the proposed algorithm to solve $\boldsymbol{A}\boldsymbol{\mathcal{X}}=\boldsymbol{B}$ in a distributed, privacy preserving fashion and to obtain $\boldsymbol{X}_i(t)$ for each ISO $i$ ($\forall i$). Thereafter, we computed the average values of the relative errors of all the elements in $\boldsymbol{X}_i(t)$ ($\forall i$) compared to $\boldsymbol{\mathcal{X}}_{real}$. The corresponding results are listed in Table \ref{case_errorAPC}. As can be observed, The relative errors are negligibly small for the chosen stopping criterion of the iterative algorithm, indicating the correctness of the proposed privacy-preserving distributed APC algorithm.
\begin{table}[h]
\setlength{\abovecaptionskip}{0pt}
	\renewcommand{\arraystretch}{1.2}
	\caption{Relative Error of Privacy-Preserving Distributed APC Algorithm}
	\label{case_errorAPC}
	\centering
	\footnotesize
	\setlength{\tabcolsep}{0.8mm}{
	\begin{tabular}{l c c c c c c c c c}
	\hline
	\bfseries ISO & 1 & 2 & 3 & 4 & 5 & 6 & 7 & 8 & 9  \\
	\hline
	\bfseries Relative Error ($10^{-7}$) & 1.69 & 1.70 & 1.61 & 1.67 & 1.65 & 1.64 & 1.80 & 1.79 & 1.81  \\
	\hline
	\end{tabular}}
\end{table}

To verify the correctness of the proposed privacy-preserving distributed PLF method, we used the centralized GMM-based PLF as benchmark. Then, we utilized the proposed PLF method for each ISO to obtain the PLF of its region. Both methods use the DLPF as the power flow model. After that, we used the Jensen--Shannon divergence (JSD) to measure the differences between the probability distributions obtained from the benchmark and proposed method. Note that the non-negative JSD between two probability distributions is bounded by 1, and smaller divergence indicates smaller differences between two probability distributions. As each region has its own probability distributions for its nodal voltages, angles, and branch flows, we computed the average and the maximal JSDs between the distributions built by the benchmark and proposed method for each region. The corresponding results are listed in Table \ref{case_error_cen}, where the bottom three rows are the maximums. Clearly, the average JSDs of all regions are negligible. Meanwhile, the maximal JSDs are all below $9.26 \times 10^{-5} \ll 1 $. Hence, the probability distributions obtained from the proposed method are basically the same as those obtained from the centralized method, verifying the correctness of the proposed distributed method.
\begin{table}[h]
\setlength{\abovecaptionskip}{0pt}
	\renewcommand{\arraystretch}{1.1}
	\caption{JSD between Centralized and Distributed PLF Methods}
	\label{case_error_cen}
	\centering
	\footnotesize
	\setlength{\tabcolsep}{1mm}{
	\begin{tabular}{l c c c c c c c c c}
	\hline
	\bfseries Region & 1 & 2 & 3 & 4 & 5 & 6 & 7 & 8 & 9  \\
	\hline
	\bfseries A Voltage ($\times 10^{-6}$)  & 2.81 & 4.97 & 0.21 & 0.02 & 0.27 & 0.16 & 3.37 & 4.48 & 0.01\\
	\bfseries V Angle ($\times 10^{-9}$) & 0.32 & 0.78 & 4.27 & 1.55 & 1.17 & 0.68 & 2.14 & 1.19 & 1.63  \\
	\bfseries E Flow ($\times 10^{-8}$) & 5.19 & 0.93 & 1.55 & 0.99 & 1.47 & 0.07 & 3.21 & 1.08 & 0.17  \\
	\hline
	\hline
	\bfseries M Voltage ($\times 10^{-5}$)  & 7.61 & 3.85 & 1.38 & 4.23 & 0.44 & 3.72 & 4.47 & 2.31 & 9.26\\
	\bfseries A Angle ($\times 10^{-7}$) & 1.08 & 1.50 & 0.90 & 0.73 & 0.93 & 0.75 & 1.41 & 1.33 & 1.34  \\
	\bfseries X Flow ($\times 10^{-6}$) & 9.02 & 7.33 & 6.70 & 3.17 & 4.50 & 7.33 & 5.08 & 8.39 & 7.12  \\
	\hline
	\end{tabular}}
\end{table}

\begin{figure}[h]
	\centering  
	\includegraphics[width=2.8in,center]{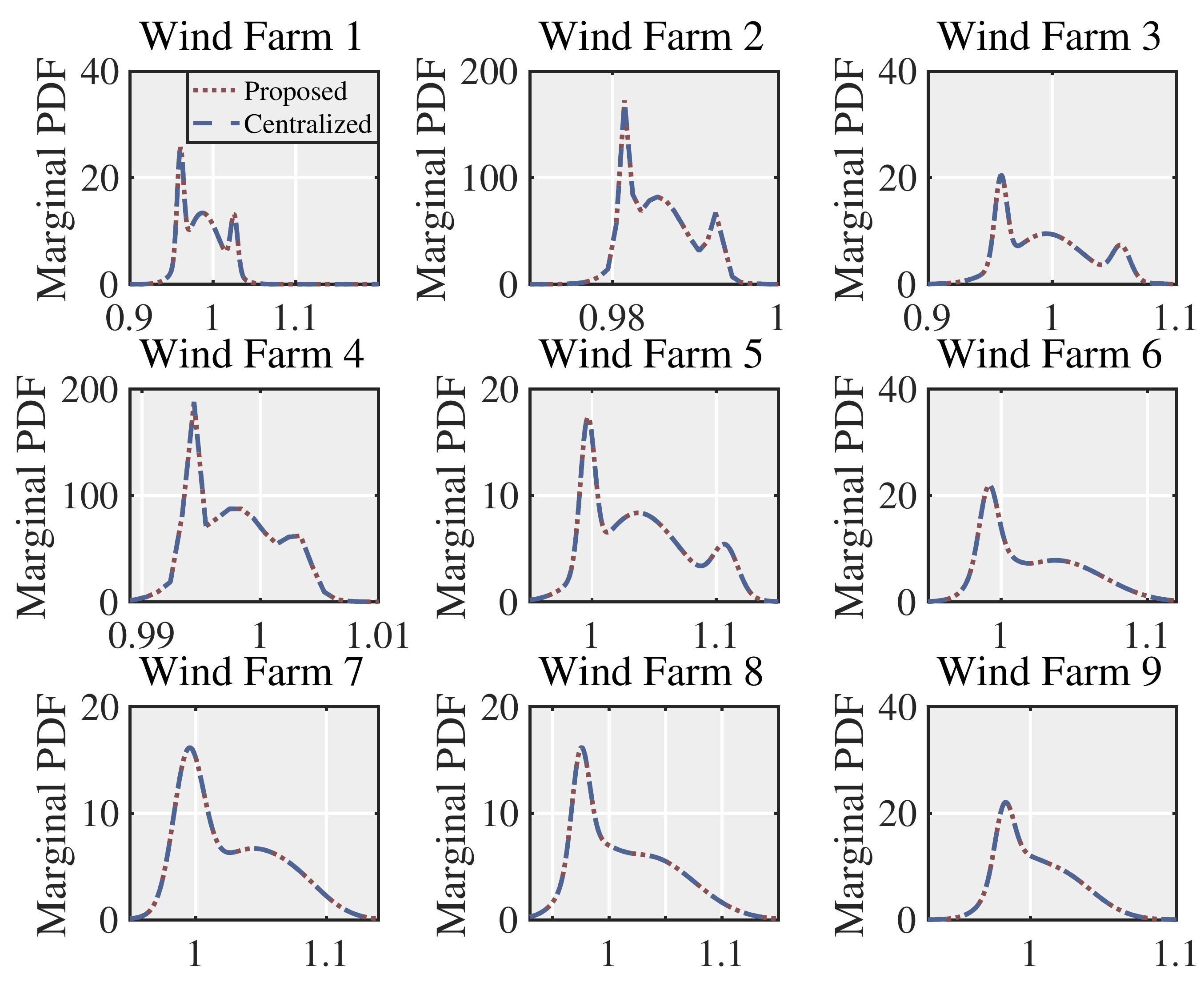}
	\caption{Voltage marginal PDFs obtained from centralized and proposed distributed PLF method.}
	\label{case_cen_dis_v}
\end{figure}

\begin{figure}[h]
	\centering  
	\includegraphics[width=2.8in,center]{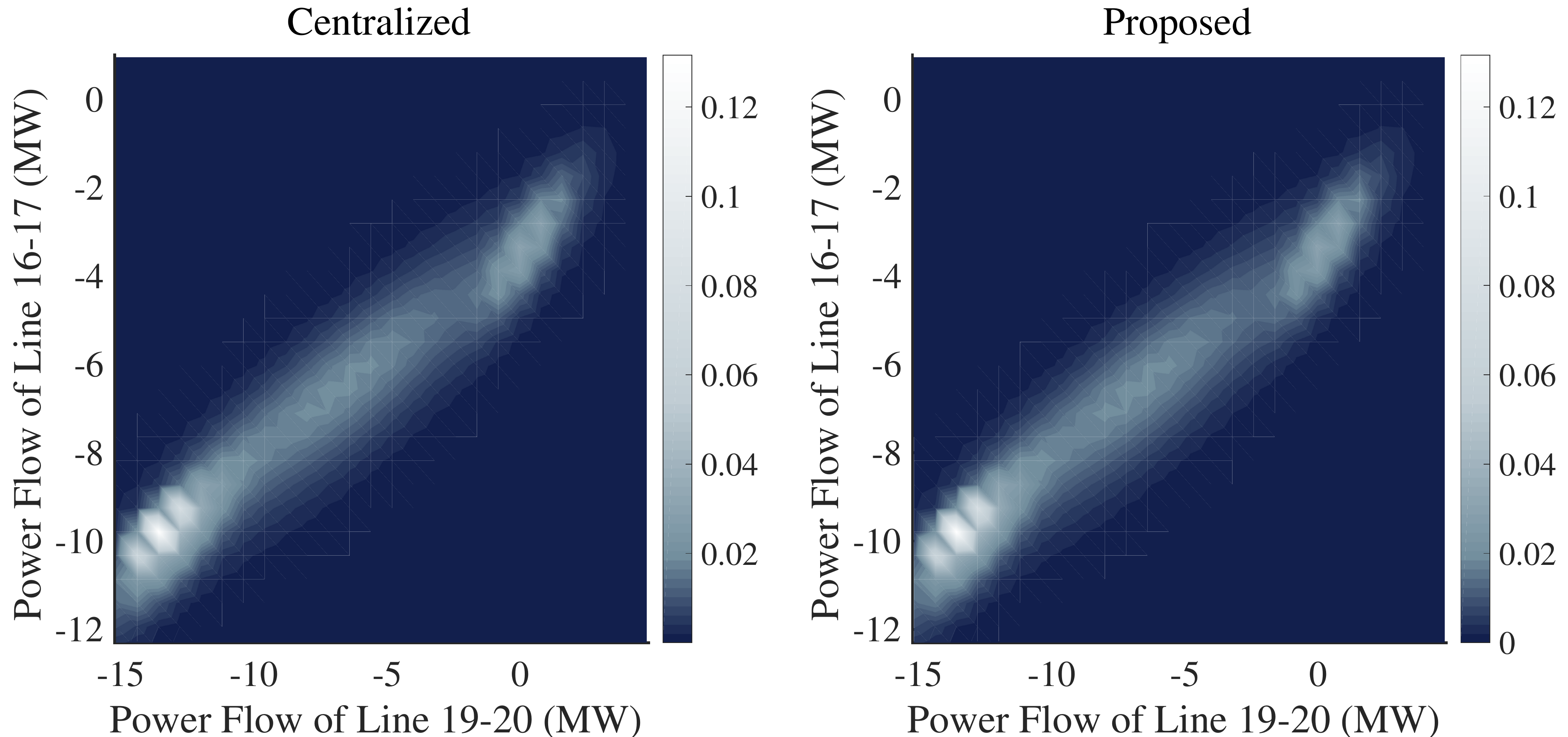}
	\caption{Branch flow joint PDFs obtained from centralized and proposed distributed PLF method.}
	\label{case_cen_dis_f}
\end{figure}

For a more detailed comparison, we illustrate the voltage probability distribution functions (PDFs) of the buses with wind farms connected in Fig. \ref{case_cen_dis_v}, because these nodal voltages have large uncertainties. In the figure, legend `Proposed' represents the PDFs obtained from the proposed method, and `Centralized' represents those obtained from the benchmark. The PDFs obtained from the benchmark and proposed method perfectly agree. We also illustrate the 2D joint PDF of two randomly chosen branch flows in Fig. \ref{case_cen_dis_f}. Again, the joint PDFs obtained from the benchmark and proposed method agree. 
\vspace{-0.5cm}

\subsection{Effectiveness Verification}
In this paper, effectiveness means that the calculation results of the analytical algorithm should be close to those of the corresponding Monte Carlo algorithm.

To verify the effectiveness of the proposed privacy-preserving distributed PLF method, we first compared it to the DLPF-based Monte Carlo method. Using this Monte Carlo method as benchmark, we calculated the average relative errors of the expected PLF values of each region obtained from the proposed method. Table \ref{case_mean_error_CDL} shows that the relative errors of each region are again negligible. Thus, the Monte Carlo and proposed methods have comparable performance. 
\begin{table}[h]
\setlength{\abovecaptionskip}{0pt}
	\renewcommand{\arraystretch}{1.1}
	\caption{Expected Value Error Using DLPF-based Monte Carlo Method as Benchmark}
	\label{case_mean_error_CDL} 
	\centering
	\footnotesize
	\setlength{\tabcolsep}{1mm}{
	\begin{tabular}{l c c c c c c c c c}
	\hline
	\bfseries Region & 1 & 2 & 3 & 4 & 5 & 6 & 7 & 8 & 9  \\
	\hline
	\bfseries Voltage ($\times 10^{-5}$)  & 1.47 & 0.19 & 0.30 & 2.26 & 0.22 & 2.27 & 3.04 & 2.07 & 1.00\\
	\bfseries Angle ($\times 10^{-4}$) & 4.76 & 3.73 & 5.11 & 3.44 & 2.39 & 1.99 & 2.67 & 2.71 & 3.43  \\
	\bfseries Flow ($\times 10^{-3}$) & 0.82 & 0.15 & 1.60 & 1.50 & 0.52 & 1.50 & 0.24 & 1.10 & 0.93  \\
	\hline
	\end{tabular}}
\end{table}

\begin{figure}[h]
	\centering  
	\includegraphics[width=2.4in,center]{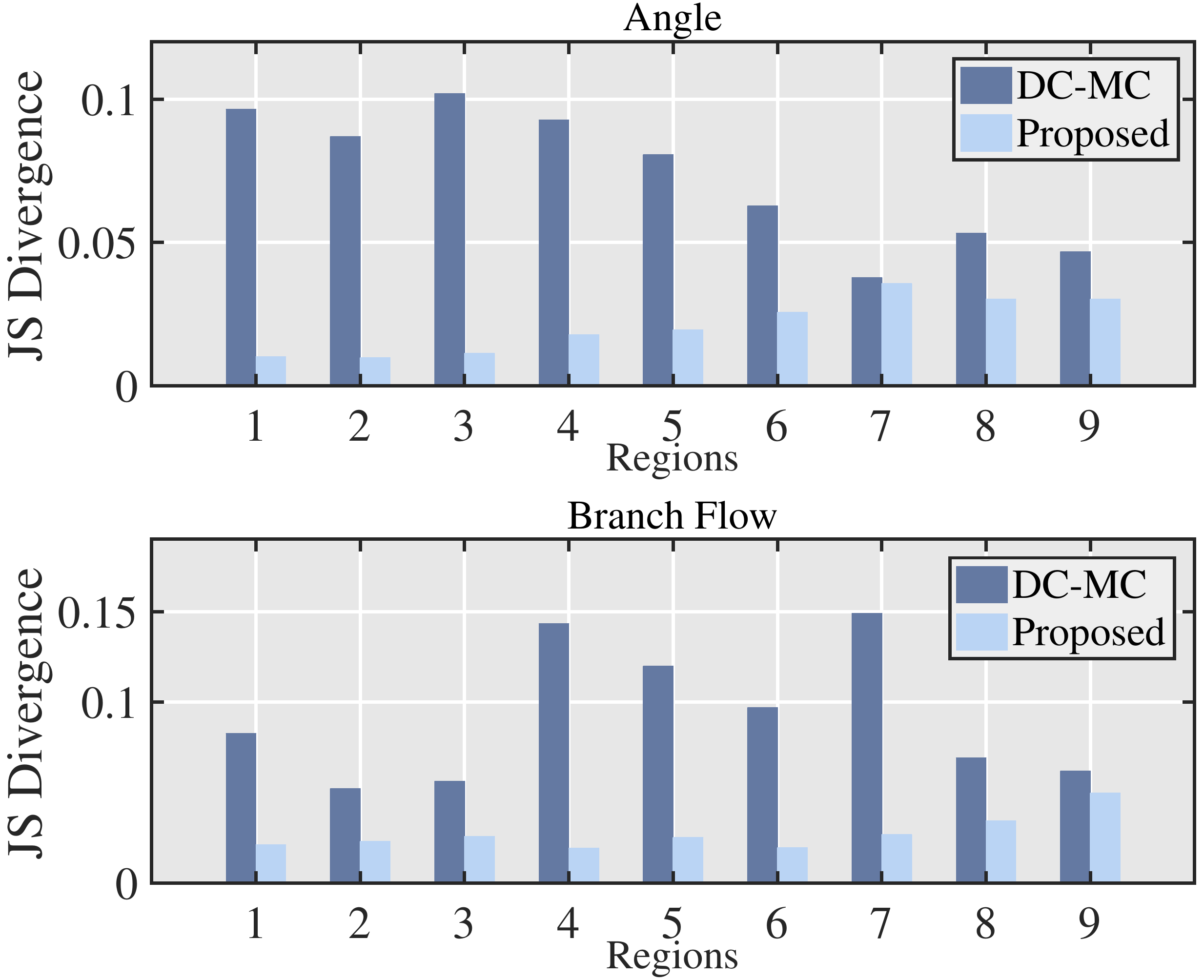}
	\caption{JSDs between the AC-based Monte Carlo method and the DC-based Monte Carlo method and the proposed method.}
	\label{case_AC_JSD}
\end{figure}

We also used the results of the AC-based Monte Carlo method as benchmark and compared the performances of the proposed method and the DC-based Monte Carlo method. The DC-based Monte Carlo method is the benchmark of the GMM-based PLF method in \cite{7574307}. The average relative errors of the expected values in each region using the above mentioned methods are listed in Table \ref{case_mean_error_AC}, with the minimal values being highlighted in bold. The relative errors of the proposed method are one to two orders of magnitude smaller than those of the DC-based Monte Carlo method. Furthermore, we used the JSD to measure the differences between the probability distributions obtained from the benchmark and the two evaluated methods. Then, we summarized the average JSD of each state and corresponding benchmark, obtaining the results depicted in Fig. \ref{case_AC_JSD}. The JSDs of the DC-based Monte Carlo method exceed 0.1, even reaching 0.15, while the JSD of the proposed method remains below $0.05$. Overall, the proposed method is superior in terms of expected value error and JSD.
\begin{table}[h]
\setlength{\abovecaptionskip}{0pt}
	\renewcommand{\arraystretch}{1.1}
	\caption{Expected Value Error Using AC-based Monte Carlo Method as Benchmark (DC-MC: DC-based Monte Carlo)}
	\label{case_mean_error_AC}
	\centering
	\footnotesize
	\setlength{\tabcolsep}{0.5mm}{
	\begin{tabular}{l c c c c c c c c c}
	\hline
	\bfseries Region & 1 & 2 & 3 & 4 & 5 & 6 & 7 & 8 & 9  \\
	\hline
	\bfseries DC-MC (Angle)  & 0.102 & 0.083 & 0.111 & 0.062 & 0.039 & 0.031 & 0.031 & 0.038 & 0.044\\
	\bfseries Proposed (Angle) & \bfseries 0.001 & \bfseries 0.002 & \bfseries 0.001 & \bfseries 0.002 & \bfseries 0.002 & \bfseries 0.003 & \bfseries 0.006 & \bfseries 0.005 & \bfseries 0.006  \\
	\hline
	\bfseries DC-MC (Flow)  & 0.085 & 0.037 & 0.191 & 0.088 & 0.085 & 0.375 & 0.227 & 0.091 & 0.050\\
	\bfseries Proposed (Flow) & \bfseries 0.003 & \bfseries 0.002 & \bfseries \bfseries 0.013 & \bfseries 0.004 & \bfseries 0.003 & \bfseries 0.016 & \bfseries 0.006 & \bfseries 0.009 & \bfseries 0.006  \\
	\hline
	\end{tabular}}
\end{table}

\begin{figure}[h]
	\centering  
	\includegraphics[width=2.1in,center]{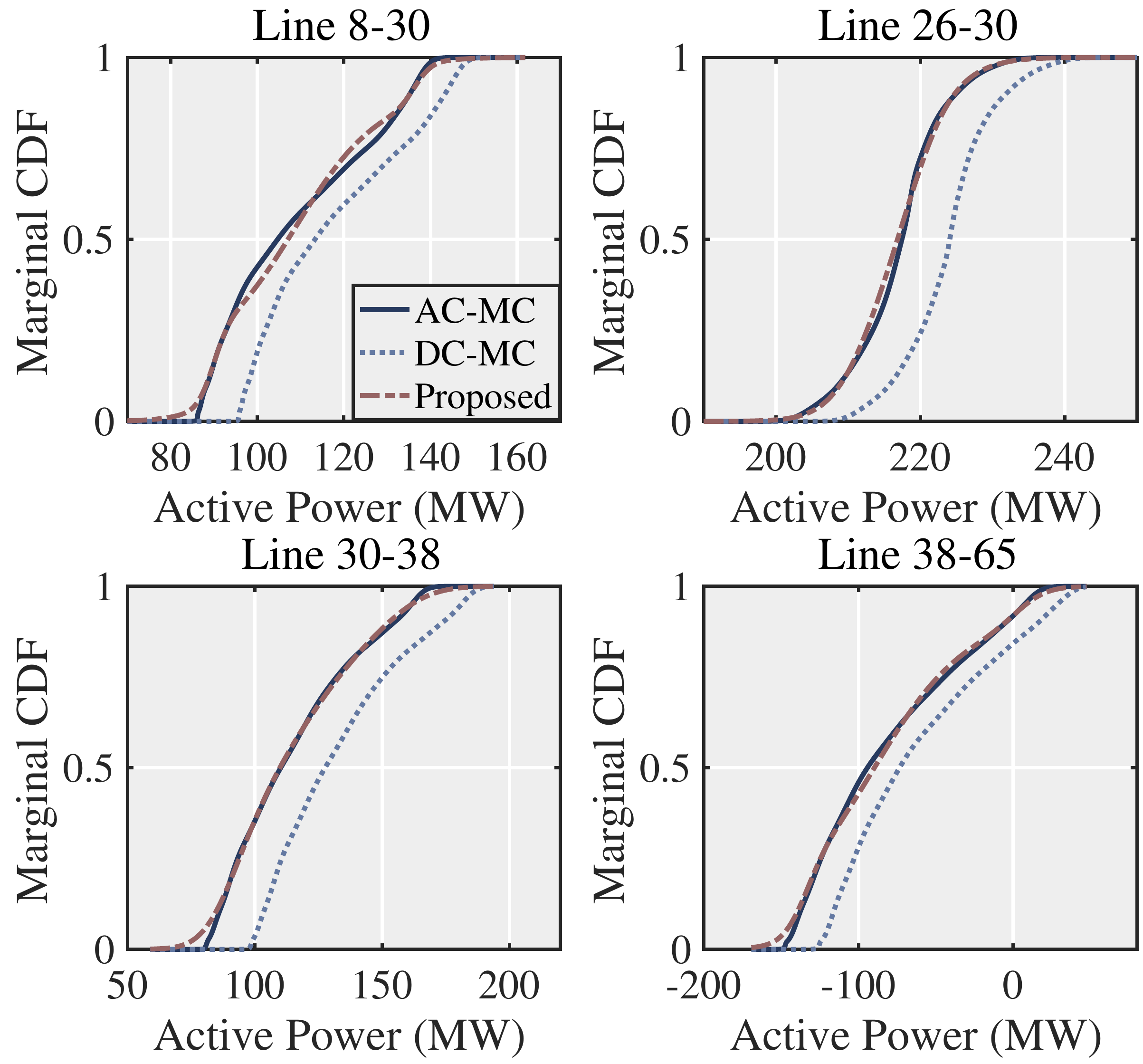}
	\caption{Branch flow marginal CDFs obtained from the AC-based Monte Carlo, DC-based Monte Carlo, and proposed methods.}
	\label{case_AC_1d_CDF}
\end{figure}

\begin{figure}[h]
	\centering  
	\includegraphics[width=3.5in,center]{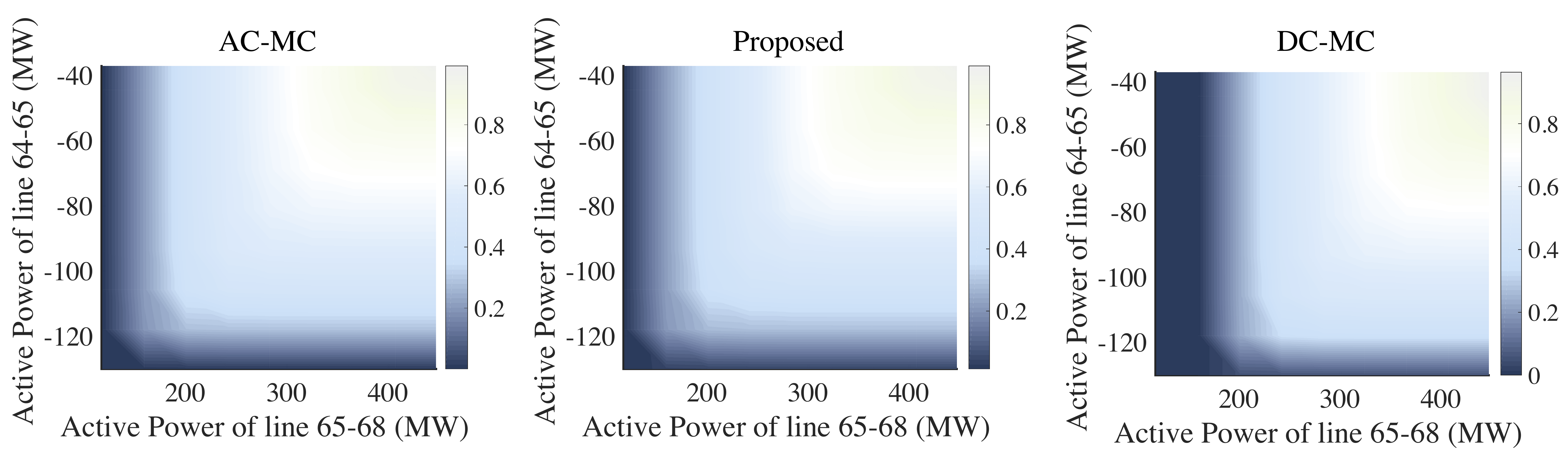}
	\caption{Branch flow joint CDFs obtained from the AC-based Monte Carlo, DC-based Monte Carlo, and proposed methods.}
	\label{case_AC_2d_CDF}
\end{figure}

For more intuitive comparisons, Fig. \ref{case_AC_1d_CDF} shows the marginal cumulative distribution functions (CDFs) of the active branch flows on a number of 345 kV transmission lines. There are clear differences between the benchmark CDF and those obtained from the DC-based Monte Carlo method. However, the CDFs obtained from the proposed method suitably agree with those obtained from the benchmark. Moreover, Fig. \ref{case_AC_2d_CDF} shows the joint CDFs of the branch flows on the 345 kV transmission lines 65-68 and 64-65. Again, the joint CDF obtained from the proposed method show better agreement with the benchmark than that obtained from the DC-based Monte Carlo method.

\subsection{Efficiency Comparison}
To verify the efficiency of the proposed privacy-preserving distributed PLF method, we measured the computational times of the evaluated methods and listed the results in Table \ref{case_time}. Note that all the Monte Carlo methods require two steps for calculation: 1) running $10^6$ power flow simulations and 2) obtaining the PLF of states in each region using the corresponding samples. The distributed method was coded in a serial structure and the computation time is given for a serial execution of the regions. 
\begin{table}[h]
\setlength{\abovecaptionskip}{0pt}
	\renewcommand{\arraystretch}{1.2}
	\caption{Computational Time of Evaluated Methods (s) (AC-MC: AC-based Monte Carlo, DC-MC: DC-based Monte Carlo)}
	\label{case_time} 
	\centering
	\footnotesize
	\setlength{\tabcolsep}{0.9mm}{
	\begin{tabular}{l c c c c c c c c c}
	\hline
	\bfseries Region & 1 & 2 & 3 & 4 & 5 & 6 & 7 & 8 & 9  \\
	\hline
	\bfseries AC-MC  & 1384 & 1177 & 1145 & 1275 & 1106 & 1199 & 1144 & 1117 & 1156\\
	\bfseries DC-MC   & 458 & 328 & 291 & 421 & 235 & 343 & 276 & 282 & 279\\
	\bfseries Proposed & 36.9 & 36.6 & 36.6 & 36.7 & 36.9 & 36.9 & 36.8 & 36.8 & 36.8  \\ 
	\bfseries Centralized & 14.52 & 14.50 & 14.50 & 14.51 & 14.50 & 14.50 & 14.50 & 14.50 & 14.50  \\
	\hline
	\end{tabular}}
\end{table}

As Table \ref{case_time} indicates, the proposed method requires about 37 seconds to obtain the PLF of a region, which is significantly faster than the Monte Carlo methods. Compared with the centralized GMM-based PLF method, the proposed method approximately costs an extra 22 seconds. This extra time could be regarded as the price of protecting ISOs' PLF and parameter information using the privacy-preserving distributed strategy.



\section{Conclusion}
For a multi-regional interconnected grid, we propose a privacy-preserving distributed PLF method to allow every regional ISO to only obtain its regional joint PLF in a fully distributed manner without revealing its parameter information to other ISOs. To this end, we first  embed the centralized GMM-based PLF into a distributed framework. In this framework, each ISO computes \eqref{Xi} and \eqref{psi n} in a fully distributed and privacy-preserving manner. We then propose a privacy-preserving distributed APC algorithm for the ISOs to calculate \eqref{Xi} and leverage the privacy-preserving AAC algorithm with fake input for the ISOs to obtain \eqref{psi n}. Combining these two algorithms, we derive the proposed privacy-preserving distributed PLF method.

Using the proposed method, each ISO only needs its own system parameters for computing the regional PLF. In addition, each ISO only needs to communicate with its neighbors, and no center for data collection is required. Moreover, no ISO can deduce the PLF and parameters of other regions despite communication. 

Case studies show that the PLF obtained from the proposed method perfectly agree with the results obtained from the centralized GMM-based PLF method. Compared to the AC-based Monte Carlo method, the accuracy of the proposed method is satisfactory, being higher than that of a benchmark used in the existing work. Moreover, the computational time of the proposed method is significantly shorter than that required by various Monte Carlo methods. The proposed method approximately costs an extra 22 seconds compared to the centralized GMM-based PLF method, which could be regarded as the price of protecting ISOs' PLF and parameter information using the privacy-preserving distributed strategy.

\bibliographystyle{IEEEtran}
\bibliography{IEEEabrv,paper}
\end{document}